\documentclass[11pt]{article}
\usepackage{amssymb}
\usepackage{graphicx}
\usepackage{setspace}
\normalbaselines

\begin{document}
\upshape

\title{\bf{ Ordered, disordered and coexistent stable vortex lattices in
$%
NbSe_{2} $ single crystals.}
}
\author{G.~Pasquini\footnote{email:pasquini@df.uba.ar} , D.~P\'{e}rez Daroca, C.~Chiliotte, G.~S.~Lozano, V.~Bekeris
\\
\small{\it{Departamento de F\'{\i}sica, FCEyN, Universidad de Buenos Aires.}}
\\
\small{\it{
Pabellon 1,Ciudad Universitaria, 1428, Buenos Aires, Argentina.}}
}
\maketitle
\begin{picture}(10,20)(10,20)
\put(14,22.6){\line(10,0){440.3}}
\end{picture}
\vspace{-0.5cm}
\begin{abstract}
\singlespacing
It is commonly accepted that the peak effect (PE) in the critical current
density of type II superconductors is a consequence of an order-disorder
transition in the vortex lattice (VL). Examination of vortex lattice
configurations (VLCs) in its vicinity requires the use of experimental
techniques that exclude current induced VL reorganization. By means of
linear $ac$ susceptibility experiments in the Campbell regime, where
vortices are forced to oscillate (harmonically) around their effective
pinning potentials, we explore quasi-static stable and metastable VLCs in $%
NbSe_{2}$ single crystals near the PE. We identify three different regions:
for $T<T_{1}(H)$, stable VLCs are maximally ordered. For $T>T_{2}(H)$
configurations are fully disordered and no metastability is observed. In the
$T_{1}<T<T_{2}$ region we find temperature dependent stable configurations
with intermediate degree of disorder, possibly associated to coexistence of
ordered and disordered lattices throughout the PE. A simple estimation of
the equilibrium proportion of ordered and disordered domains is provided.
\end{abstract}
\vspace{-0.5cm}
\begin{picture}(10,20)(10,20)
\put(14,22.6){\line(10,0){440.3}}
\end{picture}
\vspace{0.2cm}

The behavior of systems under the influence of both, thermal and quenched
disorder has been an issue of intense research over the last 50 years in
different areas of condensed matter physics. Quenched disorder can broaden
first order phase transitions inducing phenomena as phase separation (under
active study in the context of manganites for instance) and if sufficiently
strong can even change completely the properties of this system producing
the appearance of new phases (as it is the case in many glassy states) \cite%
{ps}.

In this context, the vortex lattice (VL) in superconductors provides models
systems where elastic and pinning interactions together with thermal
fluctuations compete. A fingerprints of complex behavior in this system is
the Peak Effect (PE), an anomalous non-monotonous dependence of the critical
current density $J_{_{C}}$ with both temperature and magnetic field \cite%
{pico1}. The origin and nature of this behavior are still controversial
issues.

Phenomenological pictures based in an\ order-disorder (O-D) transition from
a quasi-ordered Bragg glass (BG) \cite{giamarchi95} to a disordered phase
with proliferation of topological defects \cite{mikitik01kierfekd04} explain
a broad amount of related experimental results. However, the underlying
physics of the transition, the structural symmetry of the disordered phase
\cite{Fasano02} \texttt{and} its connection with a clean limit melting
transition \cite{Xiao04, Kokubo05} are issues that remain unsolved. While
increasing amount of experimental evidence \cite{GV06} suggests that the
nature of the PE could depend on the various materials, the\ marked
metastability and history effects are common facts\ reported both in low
\cite{lowTc,Ravikumar00,Paltiel00} and in high $T_{c}$ \cite{htc} materials.

In traditional superconductors, neutron diffraction experiments show a clear
change in the structure factor lattice at the PE indicating an O-D
transition \cite{Ling00Park03}. In $NbSe_{2}$, \ metastable coexisting
regions with different $J_{c}$\ have been directly observed \cite%
{Marchevsky01}; this fact as well as transport measurements in the Corbino
geometry \cite{Paltiel00} and experiments of magnetization assisted by a
shaking $ac$ field \cite{Ravikumar00}, suggest a first order transition
where the $ac$ magnetic field assists an equilibration process from a
supercooled disordered metastable phase to an ordered stable BG phase. \

PE and history effects are often studied by means of transport experiments.
In standard transport techniques, vortex undergo some type of
current-induced reorganization and therefore the original vortex lattice
configuration (VLC) is not accessible. To overcome this problem, Xiao et al.~%
\cite{Xiao04} have developed a sophisticated ultrafast \ transport technique
and have shown the existence of an enlarged crossing phase boundary between
ordered and disordered phases in $NbSe_{2}$ crystals. A well known and
appropriate technique to measure the pinned VL response without external
disturbances is $ac$ susceptibility restricted to the linear Campbell regime
\cite{Campbell} where a very small $ac$ field $h_{a}$ superimposed to the $%
dc $ field $H$ is applied, forcing vortices to perform small (harmonic)
oscillations inside their effective pinning potential wells.

In this letter we explore quasi-static VLCs in the Campbell regime in \ $%
NbSe_{2}$\ single crystals in the vicinity of the PE following different
thermal, magnetic and dynamical histories. Essential to\ our work, we avoid
any measurement induced VL reorganization.\ We are able to access the
corresponding \textit{stable} configuration at each temperature (field) by
applying a shaking $ac$\ field. For the first time, the pinning potential
curvature of \textit{stable} quasi-static VLCs is measured throughout the
PE, resulting in a clear identification of low temperature maximally
ordered, high temperature maximally disordered and stable configurations
with intermediate degree of disorder at intermediate temperatures. In the
last case, the \textit{stable} VLCs are separated by energy barriers,
possibly related with surfaces between coexisting ordered and disordered VL
domains. We can also observe the evolution of\ the pinning potential with
temperature in \textit{metastable} VLCs and confirm the spontaneous ordering
in cooling processes \cite{Andrei06}.

Results shown here correspond to a $NbSe_{2}$ single crystal \cite{muestras}
of approximate dimensions $(0.5\times 0.5\times 0.03)\ mm^{3}$, with $%
Tc=7.30\ K$\ (defined as the mid point of the $ac$ susceptibility linear
transition at $H=0$) and $\Delta Tc=\pm 0.02\ K$. $Ac$ susceptibility has
been measured with a home made susceptometer based in the mutual inductance
technique. All fields are parallel to the $c$ axis of the sample.

In the Campbell limit, the inductive component of $ac$\ susceptibility $\chi
^{,}$ is determined by the geometry and the dimensionless parameter $\lambda
_{R}/D$ ($\lambda _{R}$ is the real penetration depth and $D$ is a
characteristic sample dimension). In this regime the imaginary penetration
depth $\lambda _{I}<<\lambda _{R}$ . The curvature of the effective pinning
potential well is the Labusch constant $\alpha _{L}$\ that can be
numerically estimated from $\lambda _{R}=(\lambda _{L}^{2}+\phi _{0}B/4\pi
\alpha _{L})^{1/2}$, where $\lambda _{L}$ is the London penetration depth
(for references and details of the numerical procedure see Ref.\cite{Gabi99}%
).

\begin{figure}[tb]
\centering
\includegraphics [width=136mm,height=120mm]{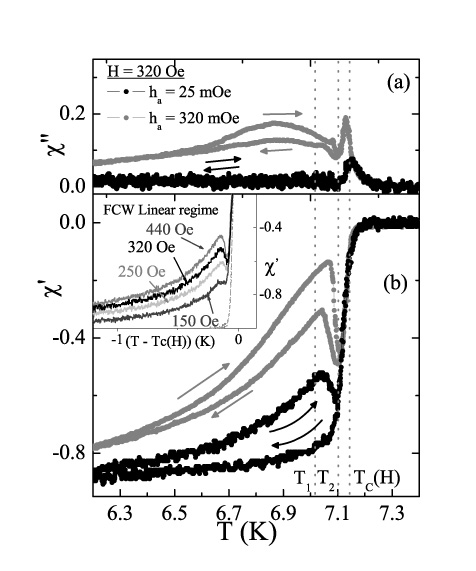}
\small{\caption{{\protect\small Typical }$\protect\chi ^{,,}(T)${\protect\small \
(panel a) and }$\protect\chi ^{,}(T)${\protect\small \ (panel b) FCC and FCW
curves in the linear (black curves) and non-linear regime (gray curves)
measured at \ }${\protect\small \emph{f}=30\ kHz}${\protect\small . Arrows
indicate the direction of temperature variation. }${\protect\small T}_{%
{\protect\small 1}}${\protect\small \ and }${\protect\small T}_{%
{\protect\small 2}}${\protect\small \ limit the }${\protect\small T}$%
{\protect\small \ regions of stable ordered and disordered phases (see
text). Inset: Linear FCW curves at various dc fields ploted as a function of
(}${\protect\small T-Tc(H)}${\protect\small ). An abrupt PE develops for }$%
{\protect\small H\geq 200\ Oe}${\protect\small . The zero dc field
transition is also shown. }}}
\label{fig1}
\end{figure}
Fig.~1 shows typical $\chi ^{,,}(T)$ (panel a) and $\chi ^{,}(T)$ (panel b)
field cool cooling (FCC) and field cool warming (FCW) curves in the linear ($%
h_{a}=0.025$ $Oe$, black) and non-linear regime ($h_{a}=0.32$ $Oe$,\ gray)
at $H=320\ Oe$. The temperatures at which $\chi ^{,}$\emph{\ }shows a
maximum or a minimum\ ($T_{p}^{on}$and $T_{p}$\ in Ref. \cite{Thakur05})
slightly depend on $h_{\emph{a}}$\ or history . Above a temperature $%
T_{2}\gtrsim T_{p}$\ FCW\ and FCC curves merge.\ Below $T_{2}$\ , $\lambda
_{I}\lesssim 0.1$\ $\lambda _{R}$ so\textsl{\ }dissipation is very small ($%
\chi ^{,,}\approx 0$) and $\chi ^{,}$ is frequency independent in the kHz
range, indicating a Campbell regime. In the non-linear regime, where
vortices are forced out of their pinning sites, both FCC and FCW curves
display PE \ whereas in the linear Campbell regime it appears only in the
warming process. As the PE is a signature of an O-D transition, its absence
in the FCC Campbell regime indicates that the VL nucleates at high T and
remains trapped in a metastable disordered and strongly pinned
configuration. On the other hand, even if $ac$ induced currents are not able
to assist the VL, the system warms with lower pinning potential (larger
penetration depth). In this framework, the hysteresis displayed in the
Campbell regime\textsl{\ }must imply some kind of spontaneous re-ordering at
low temperatures. This ordering during FCC was recently observed by using a
time resolved transport technique \cite{Andrei06}. In the inset of Fig.~2, \
$\chi ^{,}(T)$ recorded in various warming processes from different low
temperatures is shown. A lower initial temperature results in a less pinned
(more ordered) VLC. \ We have tested that the warming curves are independent
on the cooling/warming rate, and are identical if the cooling process is
performed without measurement.

This spontaneous ordering at low temperatures, can be enhanced by applying a
large shaking $ac$ field. This shaking field allows the system to explore
the free-energy landscape and to reach a more ordered and stable VLC. This
feature is illustrated in the main panel of Fig.~2. The result of the FCW
process from $4.2\ K$ shown in the inset (gray line curve in both panels )
is compared with various warming curves (symbols) recorded after shaking%
\textsl{\ }the lattice at different temperatures $T_{sh}$ (large vertical
arrows). The following process has been performed: the system has been
measured\ in the linear regime during FCC procedure and stabilized at $%
T_{sh} $. Then \ the measurement was interrupted and a sinusoidal shaking $ac
$ field of $3.2\ Oe$ - $30\ kHz$\ was applied during $30\ s$. The
measurement ($h_{a}=25\ mOe$) was resumed in a warming procedure starting
from $T_{sh}$ leading to a drastic decrease of pinning (vertical color lines
in the figure). The various symbols are hard to distinguish because all
warming procedures collapse to the same curve. This curve has remarkable
characteristics below a \ temperature $T_{1}<T_{p}^{on}$ (dotted line in the
figure): first, it is reversible; second, after any additional shaking, with
any waveform or amplitude, the system remains in this configuration \cite%
{VicPRL01}. These features constitute a first important result: below $T_{1}$%
\ the \textit{stable} VLCs (attained via smooth temperature changes) are
continuously connected in a minimum of the free-energy landscape. These are
the less pinned (more ordered) VLCs at each $T$. This picture is consistent
with a \textit{stable }ordered BG below $T_{1}$ where changes in $T$\ only
produce elastic (reversible) deformations.
\begin{figure}[tb]
\centering
\includegraphics [width=136mm,height=112mm]{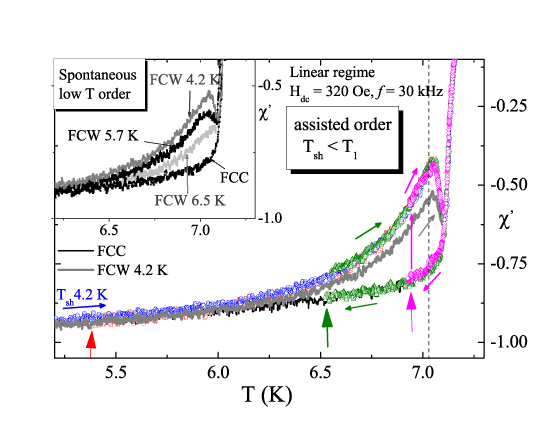}
\caption{{\protect\small (Color online). Inset: }$\protect\chi ^{,}(T)$%
{\protect\small \ recorded in various warming processes from different }$T$%
{\protect\small \ : there is a progressive spontaneous ordering in the
cooling process. Main panel: FCC and FCW (4.2 K)\ processes (gray lines)
compared with curves (various symbols) recorded before and after shaking the
VL at different temperatures }$T_{sh}${\protect\small \ (large vertical
arrows). Vertical (magenta) line remarks the huge }$\protect\chi ^{,}$%
{\protect\small \ change after shaking. }All shaken VLCs collapse to the
same reversible curve{\protect\small . Small arrows indicate the change in
temperature with time.}}
\label{fig2}
\end{figure}

The description changes completely when $T_{1}<T<T_{2}$: the more ordered
VLCs become unstable and the PE develops. Again, the system does not reach
spontaneously a stable VLC: warming (cooling) leads to metastable over
ordered (disordered) configurations\texttt{.} In every case, a large shaking
field assists the system in accessing a\textit{\ }stable VLC. The
qualitative difference observed above $T_{1}$\ is that, once a stable VLC
has been reached, any smooth temperature variation results in new metastable
states: the stable configuration at the new temperature is only reached
through shaking; consequently, no reversibility is possible. This remarkable
feature is one of the central issues of this Letter, and is illustrated in
Fig.~3 where $\chi ^{,}$ values in the linear regime corresponding to the
\textit{stable} VLCs at various $T$ are plotted in large symbols. The
complete experiment shown in Fig.~3 is the following: First a warming
process (small black dots) was performed stabilizing \ temperature\
successively at very small intervals $\Delta T_{sh}$ \cite{nota}. At each $%
T_{sh}$ (indicated by black large vertical arrows) the measurement was
interrupted and the same shaking protocol described above was performed, in
way to reach the stable VLC (open triangles). At fixed temperature,
additional\ perturbations do not modify the response and the system remains
in this \textit{stable} configuration. Then the measurement on warming was
resumed (small black dots) until the next $T_{sh}$ was reached.\ As $T$
increases the corresponding \textit{stable} VLC is more disordered.

Once $T>T_{2}$ was reached, a cooling process (small gray dots) was
performed \ and a similar protocol \ was follow\texttt{e}d, shaking the
system at the same temperatures (gray vertical arrows). Notice that $\chi
^{,}(T_{sh})$ obtained after shaking in warming (hollow triangles) and
cooling (gray circles) \ is the same (within experimental resolution). This
central result supports the stability of VLCs with intermediate order, and
rules out the possibility that they arise from a harder metastability or
slow dynamics. The more ordered (warming curve after shaking at $%
T_{sh}<T_{1} $) and the more disordered FCC curve are shown for reference in
black line.

\begin{figure}[tb]
\centering
\includegraphics [width=156mm,height=112mm]{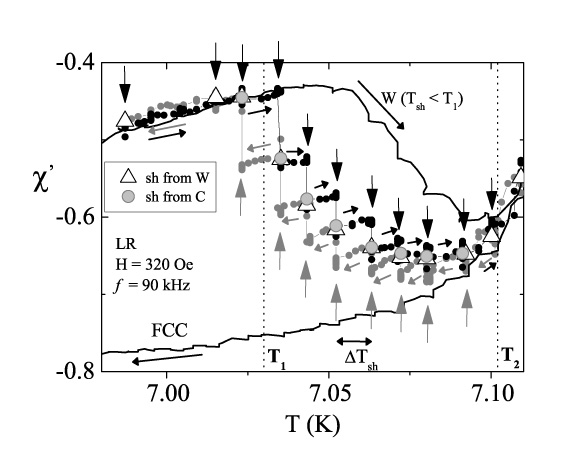}
\caption{{\protect\small Warming (W) and cooling (C) experiment \ between }$%
T_{1}${\protect\small \ and }$T_{2}$. {\protect\small Vertical down (up)
arrows identify the }$T_{sh}${\protect\small \ in the W (C) process. At each
}$T_{sh}${\protect\small \ the lattice is shaken \ to find the stable VLC \
(large open triangles in W and large gray circles in C ). The stable VLC \
warms (cools) in a metastable VLC (small dots). \ The more ordered \ and
disordered curves are also shown in black line for reference. }}
\label{fig3}
\end{figure}

\begin{figure}[tb]
\centering
\includegraphics [width=106mm]{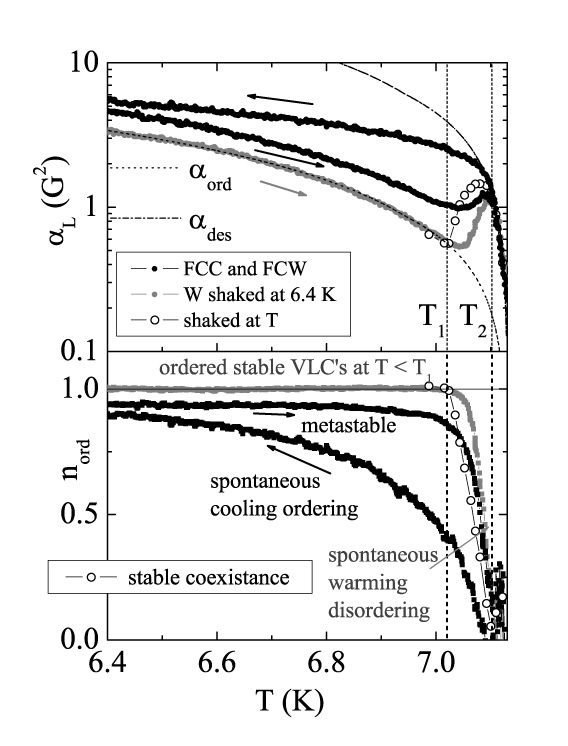}
\caption{{\protect\small (a) }$\protect\alpha _{L}(T)${\protect\small \ in
log. scale corresponding to the various stable and metastable VLCs.}
{\protect\small \ }$\protect\alpha _{ord}${\protect\small \ (}$\protect%
\alpha _{dis}${\protect\small ) is a fitting curve that \ extrapolates }$%
\protect\alpha (T)${\protect\small \ in a completely ordered (disordered)
VL. (b) Estimated proportion of ordered phase in each VLC. Arrows indicate
the change in temperature with time}}
\label{fig4}%
\end{figure}

To quantify the pinning strength on the various VLCs, the effective Labush
parameter $\alpha _{L}(T)$ corresponding to each VLC has been estimated from
the $\chi ^{,}$ values in the linear regime at $90\ kHz$ in the region where
$\lambda _{I}<<\lambda _{R}$. The result is shown in Fig.~4a. Black curves
correspond to metastable VLCs evolving in FCC and FCW processes whereas the
gray curve corresponds to ordered VLCs attained by shaking at $6.4\ K$.
Hollow circles\ show\ $\alpha _{L}$ corresponding to stable VLCs between $%
T_{1}$ and $T_{2}$ extracted from Fig.~3. It can be seen that all the $%
\alpha _{L}(T)$ curves approach at low $T$. \ At $T>T_{1}$ the ordered VLCs
(gray curve) become metastable. The $\alpha _{L}$ corresponding to the
\textit{stable} VLC increases more than five times in the PE, indicating a\
huge increase of disorder, qualitatively different to the smooth and
monotonically increasing $\alpha _{L}(T)$\ observed in YBCO crystals \cite%
{GV06, VicPRL01}. The ordered (reversible) gray curve below $T_{1}$
corresponds to a BG, free of dislocations, and the black (reversible) curve
above $T_{2}$ corresponds to a completely disordered phase. The stable
character of the VLCs with intermediate degree of disorder for $%
T_{1}<T<T_{2} $, together with additional evidence of coexisting domains in $%
NbSe_{2}$ \cite{Marchevsky01}, allows us to propose a scenario of
equilibrium phase separation. In this framework, the metastability would
arise from an exceeding proportion of ordered or disordered domains. To give
a rough estimate of the proportion of ordered phase, we proceeded in the
following way: \ we extrapolated the ordered Labusch parameter $\alpha
_{ord}(T)$ to zero at $Tc(H)$ (dotted line in Fig.~4a) and we proposed $%
\alpha _{dis}(T)=7\ \alpha _{ord}(T)$ to fit the black curve above $T_{2}$
(dash-dotted line in Fig.~4a). We then used the calculated $\alpha _{L}(T)$
to estimate the proportion of ordered phase $n_{ord}(T)=(\alpha
_{dis}(T)-\alpha _{L}(T))/(\alpha _{dis}(T)-\alpha _{ord}(T))$ corresponding
to each VLC. The result is shown in Fig.~4b. \

We arrive then to the following scenario. There are three different
temperature regions along the PE: for $T<T_{1}(H)$, stable VLCs are
maximally ordered and accessible by the application of a shaking $ac$ field.
We identify these ordered stable VLCs as a BG, continuously connected by
elastic deformations in a minimum of the free-energy landscape. The robust
metaestable states below $T_{1}$ show spontaneous ordering at low $T$ as
elastic interactions overcome pinning forces, although remain separated from
the BG by a finite barrier in all the experiemntal range, down to $4.2K$.
For $T>T_{2}(H)$ configurations are fully disordered and no metastability is
observed. We identify $T_{2}(H)$\ with $Ts(H)$, the spinodal line in Ref
\cite{Thakur05}.The evolution from the ordered BG to the high temperature
disordered phase, in the small $T_{1}<T<T_{2}$ region, occurs by a gradual
increase in the proportion of disorder. The equilibrium VLCs are only
accessed from over-ordered or over-disordered configurations by a shaking $%
ac $ field, that provides sufficient energy for plastic (irreversible)
deformations to access the stable VLC. Plastic changes imply the creation or
annihilation of dislocations, probably related to the movement of domain
walls in a scenario of equilibrium phase separation. In conclusion, by
measuring for the first time the pinning potential curvature of stable
quasi-static VLCs, avoiding vortex lattice reorganization, we are able to
present a consistent picture that describes the vortex lattice physics in
the PE region in $NbSe_{2}$ single crystals.

{\bf Acknowledgments}
We thank E. Fradkin, G. Borsi and C. Acha for discussions. This
work was partially supported by UBACyT X142, UBACyT X200 and CONICET.


\begin{thebibliography}{99}
\bibitem{nota} This $\Delta T_{sh}$\ is very well determined by the smallest
step in the setpoint of the temperature controller.

\newpage
\end{thebibliography}
\end{document}